\documentclass[a4paper,11pt]{article}
\usepackage{jcappub}
\usepackage{bm}

\newcommand{\dd}{\mathrm d}
\newcommand{\ii}{\mathrm i}
\newcommand{\kvec}{\mathbf k}
\newcommand{\vvec}{\mathbf v}
\newcommand{\Vvec}{\mathbf V}
\newcommand{\Xvec}{\mathbf X}
\newcommand{\Fvec}{\mathbf F}
\newcommand{\xvec}{\mathbf x}
\newcommand{\phip}{\Phi_{\mathrm p}}

\newcommand{\Ch}{\mathrm{Ch}}

\newcommand{\dint}[1]{\int\!\frac{\dd^{3}#1}{(2\pi)^3}}

\title{Finite-wavelength corrections to dynamical friction in fuzzy dark matter: a causal linear-response approach}

\author[a,b]{Yu Wang}
\author[b,*]{Meilin Liu,}
\author[a]{Haiguang Xu}
\author[b]{Shoupan Liu}

\affiliation[a]{School of Physics and Astronomy, Shanghai Jiao Tong University, Shanghai 200240, China}

\affiliation[b]{School of Aeronautics and Astronautics, Shanghai Jiao Tong University, Shanghai 200240, China}

\emailAdd{meilin.liu@sjtu.edu.cn}

\abstract{
A massive body moving through fuzzy dark matter (FDM) generates a density wake whose gravitational backreaction produces dynamical friction. We investigate how finite-wavelength effects modify this force using causal linear-response theory and the statistical Wigner formulation. For a homogeneous Maxwellian background and general isotropic finite-size perturbers, we derive an analytic representation of the gravitational drag and recover the classical Chandrasekhar result under matched regularization conditions. In the controlled semiclassical regime, the leading finite-wavelength correction is quadratic in $\hbar/m_\chi$ and factorizes into a velocity-dependent function and a ratio of spectral moments characterizing the source structure. Remarkably, this correction reduces the drag magnitude independently of the detailed radial source profile within the adopted Maxwellian model. Gaussian and exponential sources illustrate this result. We further establish a controlled remainder bound and clarify the nonuniform behavior of the point-source limit. These findings identify the physical conditions under which wave effects modify classical dynamical friction.
}

\keywords{Fuzzy dark matter, Dynamical friction, Causal linear response, Finite-wavelength effects}

\begin{document}
\maketitle
\flushbottom

\section{Introduction}
\label{sec:intro}

A massive object moving through a gravitating medium excites a density wake whose gravitational field reacts back on the object, producing dynamical friction. Chandrasekhar's theory provides the standard description of this effect in a homogeneous collisionless background \cite{Chandrasekhar1943,BinneyTremaine2008}. In the Maxwellian approximation, the drag depends on the background density, velocity dispersion, and perturber velocity, while its application requires prescriptions for gravitational cutoffs and the spatial extent of the wake. More general treatments have emphasized resonant responses, temporal nonlocality, and the distinction between conservative and dissipative effects \cite{TremaineWeinberg1984,NelsonTremaine1999}. Related studies of gaseous media have also demonstrated the importance of wake propagation and evolution \cite{Ostriker1999}. These developments establish dynamical friction as a gravitational response of the surrounding medium rather than a purely local interaction.

Fuzzy dark matter (FDM), consisting of ultralight bosonic fields with astrophysically relevant de Broglie wavelengths, provides a natural setting in which wave effects may modify gravitational dynamics. Its wave properties can suppress small-scale structure and support characteristic gravitational configurations \cite{Hu2000,Hui2017,Marsh2016,Ferreira2021,Hui2021}. In the highly occupied nonrelativistic regime, FDM is described by the Schr\"odinger--Poisson equations. Numerical studies have revealed solitonic cores, interference patterns, and time-dependent density fluctuations \cite{Schive2014Nature,Schive2014PRL,Schwabe2016,Mocz2017,Veltmaat2018}. These features motivate a wave-based description of dynamical friction, particularly when the relevant wavelength becomes comparable to the spatial scales of the gravitational perturbation.

The connection between wave mechanics and classical collisionless dynamics is central to this problem. Early Schr\"odinger representations of cosmological matter \cite{WidrowKaiser1993} and subsequent studies of the Schr\"odinger--Poisson--Vlasov correspondence \cite{Mocz2018} established important links between the two descriptions. Nevertheless, a coherent cold wave and a statistical wave ensemble with finite velocity dispersion are physically distinct. The latter provides a natural framework for comparison with Maxwellian dynamical friction, while wave interference can additionally generate stochastic density fluctuations and gravitational relaxation \cite{BarOr2019}. A consistent calculation must therefore distinguish the mean gravitational response from fluctuation-induced effects.

Dynamical friction in FDM has been investigated in several complementary settings. Lancaster \emph{et al.} studied pointlike and extended perturbers, finite velocity dispersion, and comparisons between analytical approximations and nonlinear simulations \cite{Lancaster2020}. Buehler and Desjacques examined circular motion, gravitational wakes, and finite-time response \cite{Buehler2023}. More recently, Mitra and Wilson-Gerow investigated gravitational backreaction in coherent and stochastic ultralight-scalar environments, including binary systems and wavelength-dependent effects \cite{MitraWilsonGerow2026}. Building on these developments, we investigate the systematic structure of leading finite-wavelength corrections for general isotropic finite-size perturbers. Our central questions concern their dependence on source structure, the sign of the leading correction, and the conditions required for a controlled semiclassical expansion.

The Wigner phase-space formulation provides a natural connection between statistical wave dynamics and classical kinetic theory \cite{Wigner1932,Moyal1949,ManfrediHaas2001,Haas2011}. Unlike the classical Vlasov susceptibility, which involves derivatives of the velocity distribution, the Wigner response contains symmetric finite shifts determined by the wavelength. Its semiclassical expansion therefore generates systematic finite-wavelength corrections. However, an expansion at fixed wave number need not remain valid after integration over all gravitational scales, making the ultraviolet structure of the source essential. The corresponding resonant response is closely related to established collisionless kinetic theory \cite{Landau1946}.

Gravitational backreaction is inherently causal because the induced wake depends on the previous motion of the perturber. Kubo response theory and the Schwinger--Keldysh closed-time-path formalism provide systematic descriptions of retarded dynamical responses \cite{Kubo1957,Schwinger1961,Keldysh1965}. Influence-functional methods further distinguish deterministic environmental reactions from statistical fluctuations \cite{FeynmanVernon1963,CaldeiraLeggett1983,Chou1985,RammerSmith1986,Kamenev2023}. Related approaches have been applied to coherent and corpuscular components of bosonic dark matter, including dissipative and stochastic dynamics \cite{Proukakis2023,Proukakis2025}. Here we employ the influence-functional framework to establish the causal gravitational reaction force and the Wigner representation to determine the statistical density susceptibility.

We consider a homogeneous, isotropic Maxwellian FDM background and a finite-size gravitational perturber moving at constant velocity. The source is characterized by a normalized spherical Fourier form factor $W(k)$, allowing a general class of spatial mass distributions. We work within the bare linear-response approximation, neglecting collective self-gravitating dressing and nonlinear gravitational focusing. Identical source profiles and infrared cutoffs are employed in the wave and classical calculations, ensuring a consistent comparison. Within this framework, the complete finite-wavelength drag can be expressed through the absorptive part of the retarded Wigner susceptibility. For a Maxwellian background, the angular integration is analytic, reducing the force to a one-dimensional wave-number integral while retaining the full wavelength dependence of the linear response. The classical Chandrasekhar result is recovered in the matched semiclassical limit.

The principal result of this work is a general analytical expression for the leading finite-wavelength correction to dynamical friction for isotropic finite-size perturbers. We show that the relative quadratic correction factorizes into a velocity-dependent function and a ratio of spectral moments determined by $|W(k)|^2$. Within the Maxwellian model, the angular response fixes the direction of the correction, whereas the source structure determines its magnitude. For sources satisfying the required convergence conditions, the leading correction acts opposite to the classical drag and therefore reduces its magnitude in the controlled semiclassical regime. We demonstrate this result using Gaussian and exponential mass distributions and establish an explicit fourth-order remainder bound involving a higher spectral moment. The analysis further identifies the nonuniformity of the point-source limit and distinguishes the validity of the semiclassical expansion from the ultraviolet convergence of the complete wave response.

These results establish a systematic spectral-moment description of finite-wavelength gravitational drag, connecting the statistical properties of the FDM background with the spatial structure of the perturbing source. The resulting factorization, definite leading-order sign within the Maxwellian model, and controlled remainder provide analytical tools for investigating wave-induced modifications of dynamical friction.

The paper is organized as follows. Section~\ref{sec:sp}
introduces the Schr\"odinger--Poisson system and cold-wave
response. Section~\ref{sec:ctp} develops the causal
influence-functional formulation. Section~\ref{sec:wigner}
derives the statistical Wigner susceptibility, while
Sec.~\ref{sec:drag} presents the finite-wavelength drag
and its matched classical limit. Section~\ref{sec:correction}
establishes the leading correction for general isotropic
sources and the corresponding remainder estimates.
Section~\ref{sec:limits} discusses the physical regime
of validity and the point-source limit. Finally,
Sec.~\ref{sec:conclusion} summarizes the main findings
and discusses their physical implications and possible
extensions. Technical derivations and consistency checks
are presented in the Appendices.

\section{Schr\"odinger--Poisson dynamics and the cold-wave reference}
\label{sec:sp}

\subsection{Field action and gravitational source}

Let $m_\chi$ be the boson mass and $\Psi$ the nonrelativistic field,
normalized so that $\rho=m_\chi|\Psi|^2$. Writing $\varphi$ for the
self-gravitating Newtonian potential and $\phip$ for an imposed potential,
consider the action
\begin{equation}
\begin{aligned}
S[\Psi,\varphi;\phip]=\int \dd t\,\dd^3x\,\Bigg\{
&\frac{\ii\hbar}{2}\left(\Psi^*\dot\Psi-\dot\Psi^*\Psi\right)
-\frac{\hbar^2}{2m_\chi}|\bm\nabla\Psi|^2\\
&-m_\chi(\varphi+\phip+\Phi_0)|\Psi|^2
-\frac{|\bm\nabla\varphi|^2}{8\pi G}\Bigg\}.
\end{aligned}
\label{eq:SPaction}
\end{equation}
The prescribed background potential $\Phi_0$ is included only when a
specific background equilibrium is considered. Variation with respect to
$\Psi^*$ and $\varphi$ gives
\begin{align}
\ii\hbar\partial_t\Psi&=-\frac{\hbar^2}{2m_\chi}\nabla^2\Psi
+m_\chi(\Phi_0+\varphi+\phip)\Psi,\label{eq:SP}\quad\\
\nabla^2\varphi&=4\pi Gm_\chi|\Psi|^2.\label{eq:Poisson}
\end{align}
For a point perturber of mass $\mu$ at $\Xvec(t)$,
\begin{equation}
\nabla^2\phip=4\pi G\mu\,\delta^{(3)}[\xvec-\Xvec(t)].
\label{eq:source}
\end{equation}
We subtract any uniform background contribution to $\varphi$ when
linearizing. An infinite homogeneous self-gravitating background has a
Jeans-unstable long-wavelength sector; our steady-state drag calculations
below neglect collective wake self-gravity unless explicitly stated.

\subsection{Madelung form and linear density waves}

Introduce an amplitude and velocity potential,
\begin{equation}
\Psi=\sqrt{\rho/m_\chi}\,\exp(\ii m_\chi\vartheta/\hbar),
\qquad \mathbf u=\bm\nabla\vartheta.
\end{equation}
The field equation yields the continuity and Euler equations,
\begin{align}
\partial_t\rho+\bm\nabla\!\cdot(\rho\mathbf u)&=0,\label{eq:cont}\\
\partial_t\mathbf u+(\mathbf u\!\cdot\!\bm\nabla)\mathbf u
&=-\bm\nabla(\Phi_0+\varphi+\phip+Q_{\rm w}),\label{eq:euler}
\end{align}
where the wave-gradient potential is
\begin{equation}
Q_{\rm w}=-\frac{\hbar^2}{2m_\chi^2}
\frac{\nabla^2\sqrt\rho}{\sqrt\rho}.
\label{eq:Q}
\end{equation}
The term is often called ``quantum pressure,'' although in the
high-occupation regime it represents the dispersive gradient energy of a
classical wave field.

For a locally uniform background, write
$\rho=\rho_0+n$ and $\mathbf u=\mathbf u_0+\delta\mathbf u$.
The wave-gradient potential linearizes to
\begin{equation}
\delta Q_{\rm w}
=-\frac{\hbar^2}{4m_\chi^2\rho_0}\nabla^2n.
\end{equation}
With $D_0=\partial_t+\mathbf u_0\cdot\bm\nabla$, eliminating the
velocity perturbation gives the sourced density equation
\begin{equation}
\left(D_0^2+\frac{\hbar^2}{4m_\chi^2}\nabla^4
-4\pi G\rho_0\right)n=\rho_0\nabla^2\phip.
\label{eq:denswave}
\end{equation}
For the Fourier convention $A(\xvec,t)=\int\!\dd^3k\,\dd\omega
\,A(\kvec,\omega)e^{\ii(\kvec\cdot\xvec-\omega t)}/(2\pi)^4$,
we define $n=\chi^R_{\rm cold}\phip$ and obtain
\begin{equation}
\chi^R_{\rm cold}(\kvec,\omega)=
\frac{\rho_0k^2}{[\omega-\kvec\cdot\mathbf u_0+\ii0^+]^2
-\hbar^2k^4/(4m_\chi^2)+4\pi G\rho_0}.
\label{eq:coldchi}
\end{equation}
This formula is a useful consistency test of the action and the Madelung
linearization, not a substitute for a statistical-background susceptibility.
The real-frequency retarded prescription in Eq.~\eqref{eq:coldchi}
assumes a stable Fourier mode. With self-gravity retained, modes with
$k<k_J=(16\pi G\rho_0 m_\chi^2/\hbar^2)^{1/4}$ are Jeans unstable;
their initial-value response must instead be defined by a causal
Laplace transform with its contour above the growing poles. We do not
assign a steady drag to these unstable homogeneous modes. When
collective wake self-gravity is neglected, the last term in the
denominator is omitted.

\section{Causal response and the influence functional}
\label{sec:ctp}

The force on a gravitational perturber is a real-time expectation value.
To specify causality one must use an in-in formulation rather than identify
a time-ordered propagator with a retarded susceptibility. In the CTP
formalism the wave field is duplicated on a forward and backward branch,
$\Psi_+$ and $\Psi_-$, and
\begin{equation}
Z[\Phi_+,\Phi_-]=\int\!\mathcal D\Psi_+\mathcal D\Psi_-\,
\rho_{\rm init}\,\exp\left[\frac{\ii}{\hbar}
(S[\Psi_+,\Phi_+]-S[\Psi_-,\Phi_-])\right],
\label{eq:CTP}
\end{equation}
with the branch fields sewn together at the final time. The initial
statistical state $\rho_{\rm init}$ specifies the background occupation
numbers and correlations. Introduce classical and difference sources,
$\Phi_c=(\Phi_++\Phi_-)/2$ and $\Phi_\Delta=\Phi_+-\Phi_-$.

Expansion of the influence action to second order in the sources has the
standard causal structure
\begin{equation}
S_{\rm IF}^{(2)}=-\int\dd^4x\,\dd^4x'\,
\Phi_\Delta(x)\chi_\rho^R(x,x')\Phi_c(x')
+\frac{\ii}{2}\int\dd^4x\,\dd^4x'\,
\Phi_\Delta(x)N(x,x')\Phi_\Delta(x') .
\label{eq:IF}
\end{equation}
The mean linear response is controlled by $\chi_\rho^R$; $N$ encodes
statistical density fluctuations. The linear background term
$-\int\dd^4x\,\rho_0\Phi_\Delta(x)$ has been subtracted from
Eq.~\eqref{eq:IF}, since it produces no density \emph{wake}.
In the microscopic description one may define the symmetric noise kernel as
$N(x,x')=\langle\{\delta\hat\rho(x),\delta\hat\rho(x')\}\rangle_0/(2\hbar)$,
with the precise normalization following the chosen CTP conventions.
The imaginary part of the influence action is therefore distinct from the
\emph{absorptive} part of the retarded response: the former represents
background fluctuations, whereas the latter determines the mean drag.
With the source coupling $-\rho\Phi$ in the Lagrangian, our sign convention is
\begin{equation}
\chi_\rho^R(x,x')=-\frac{\ii}{\hbar}\Theta(t-t')
\langle[\widehat\rho(x),\widehat\rho(x')]\rangle_0,
\label{eq:Kubo}
\end{equation}
so that
\begin{equation}
\delta\rho(x)=\int\dd^4x'\,\chi_\rho^R(x,x')\phip(x').
\label{eq:linresp}
\end{equation}
The commutator form applies in the microscopic field-theory description;
in the macroscopic high-occupancy limit its response has a corresponding
classical Poisson-bracket representation.
For a perturber with mass profile $\rho_p(\xvec-\Xvec)$, varying the physical
source trajectory in the \emph{same} influence functional produces
\begin{equation}
\Fvec_{\rm wake}(t)=-\int\dd^3x\,\rho_p(\xvec-\Xvec(t))
\bm\nabla\delta\Phi_{\rm w}(\xvec,t).
\label{eq:extendedwakeforce}
\end{equation}
The wake here is the \emph{induced} response, not the bare gravitational
self-field of the source. Because $\delta\rho$ is retarded in the applied
potential, Eq.~\eqref{eq:extendedwakeforce} generally depends on the
source history. In the stationary translation-invariant limit, this
nonlocal force admits the simpler spectral representation used below.

The induced wake potential obeys
\begin{equation}
\nabla^2\delta\Phi_{\rm w}=4\pi G\,\delta\rho.
\end{equation}
The mean reaction force for a point perturber is
\begin{equation}
\Fvec_{\rm DF}(t)=-\mu\bm\nabla\delta\Phi_{\rm w}
\big|_{\xvec=\Xvec(t)}.
\label{eq:pointforce}
\end{equation}
Consequently the force is generally a functional of the earlier source
history, not an instantaneous local drag law. This nonlocality is a
standard property of retarded response; it is not claimed here as a new
phenomenon.

\section{Statistical-wave Wigner response}\label{sec:wigner}

To compare FDM with classical collisionless friction without changing the
background velocity distribution, introduce a homogeneous statistical wave
state with mass phase-space density $F_0(\vvec)$ normalized by
\begin{equation}
\int\dd^3v\,F_0(\vvec)=\rho_0,
\qquad \eta\equiv\frac{\hbar}{m_\chi}.
\label{eq:eta}
\end{equation}
The Wigner representation of the equal-time two-point function is
\begin{align}
W(\xvec,\mathbf p)&=\frac{1}{(2\pi\hbar)^3}
\int\dd^3s\,e^{-\ii\mathbf p\cdot\mathbf s/\hbar}
\nonumber\\[-2pt]
&\quad\times\langle\Psi^*(\xvec-\mathbf s/2)
\Psi(\xvec+\mathbf s/2)\rangle.
\label{eq:Wigner}
\end{align}
The background $F_0$ is obtained from a corresponding mass-normalized
Wigner function. A Gaussian statistical field state with this two-point
function can also be specified as the initial state in Eq.~\eqref{eq:CTP}.
Let $n_0(\mathbf p)$ denote the occupation of a Gaussian, homogeneous
statistical wave state. The density-response bubble from the CTP theory is
\begin{equation}
\begin{aligned}
\chi_\eta^R(\kvec,\omega)=\frac{m_\chi^2}{\hbar}
\int\frac{\dd^3p}{(2\pi\hbar)^3}
\frac{n_0(\mathbf p-\hbar\kvec/2)
-n_0(\mathbf p+\hbar\kvec/2)}
{\omega-\kvec\cdot\mathbf p/m_\chi+\ii0^+}.
\end{aligned}
\label{eq:occupationbubble}
\end{equation}
Here $n_0(\mathbf p)$ is the dimensionless occupation per momentum
state, normalized so that
$\rho_0=m_\chi\int\dd^3p\,n_0(\mathbf p)/(2\pi\hbar)^3$.
The mass-normalized distribution is
$F_0(\vvec)=m_\chi^4 n_0(m_\chi\vvec)/(2\pi\hbar)^3$.
Equation~\eqref{eq:occupationbubble} is the \emph{bare} density bubble
of a Gaussian, freely propagating statistical ensemble; interactions
and collective self-gravity are not included in that bubble.
With this convention the CTP density bubble and the linearized Wigner
equation give the same bare retarded susceptibility. Importantly the Maxwellian background here
is a \emph{statistical state}; it is not obtained by replacing the velocity
of the coherent cold Madelung solution by a dispersion parameter.
The retarded density response is
\begin{equation}
\chi_\eta^R(\kvec,\omega)=-\frac{1}{\eta}
\int\dd^3v\,
\frac{F_0(\vvec+\eta\kvec/2)-F_0(\vvec-\eta\kvec/2)}
{\omega-\kvec\cdot\vvec+\ii0^+}.
\label{eq:Wchi}
\end{equation}
This finite difference rather than a local velocity derivative is the
specific wave feature of the statistical response. It can be derived
by linearizing the Wigner kinetic equation or by evaluating the retarded
density-density bubble of the Gaussian statistical field. The coherent
cold-wave response follows from $F_0(\vvec)=\rho_0\delta^{(3)}(\vvec)$
when collective self-gravity is neglected.

At fixed $\kvec$ for a sufficiently smooth $F_0$, let
$D_\kvec=\kvec\cdot\bm\nabla_v$. The symmetric difference yields
\begin{equation}
\frac{F_0(\vvec+\eta\kvec/2)-F_0(\vvec-\eta\kvec/2)}{\eta}
=D_\kvec F_0+\frac{\eta^2}{24}D_\kvec^3F_0
+O(\eta^4),
\label{eq:Wexp}
\end{equation}
which implies
\begin{align}
\chi_\eta^R&=\chi_0^R+\eta^2\chi_2^R+O(\eta^4),\\
\chi_0^R&=-\int\dd^3v\,
\frac{D_\kvec F_0}{\omega-\kvec\cdot\vvec+\ii0^+},\label{eq:Vlasov}\\
\chi_2^R&=-\frac{1}{24}\int\dd^3v\,
\frac{D_\kvec^3F_0}{\omega-\kvec\cdot\vvec+\ii0^+}.
\label{eq:chi2}
\end{align}
Thus $\chi_\eta^R\to\chi_0^R$, the Vlasov susceptibility, in the
controlled short-wavelength limit. Importantly the expansion is initially
at fixed wave number; its interchange with the subsequent drag integral
requires an additional argument.

If wake self-gravity is retained, the response to the external potential
is instead dressed as
\begin{equation}
\chi_{\rm dressed}^R=
\frac{\chi_\eta^R}{1+(4\pi G/k^2)\chi_\eta^R}.
\label{eq:dressed}
\end{equation}
Our analytic drag calculation below uses the bare $\chi_\eta^R$.
This neglect of collective dressing is an independent approximation,
not a consequence of merely taking the source mass to be small. It is
controlled only when $|(4\pi G/k^2)\chi_\eta^R|$ is small over the
wave-number and frequency regions responsible for the measured drag,
and when the reference state is stable. Near collective poles, or on
unstable modes, the bare result need not approximate the dressed one.

\section{Complete finite-wavelength drag and the classical limit}
\label{sec:drag}

\subsection{A uniformly moving finite source}

Let a perturber move with fixed velocity $\Vvec$, with the coordinate
origin chosen such that $\Xvec(t)=\Vvec t$, and have an isotropic
mass form factor $W_b(k)$,
\begin{equation}
\phip(\kvec,\omega)=
-\frac{4\pi G\mu}{k^2}W_b(k)
(2\pi)\delta(\omega-\kvec\cdot\Vvec).
\label{eq:perturb}
\end{equation}
Both the excitation of the background and the reaction force on the
extended source involve this form factor. Using Poisson's equation and
Eq.~\eqref{eq:Wchi}, the steady dissipative force is
\begin{equation}
\Fvec_\eta=(4\pi G)^2\mu^2\dint{k}
\frac{\kvec}{k^4}|W_b(k)|^2
\operatorname{Im}\chi_\eta^R(\kvec,\kvec\cdot\Vvec).
\label{eq:masterforce}
\end{equation}
For a stable isotropic passive background, the angularly averaged force
points opposite to $\Vvec$. Equation~\eqref{eq:masterforce} makes it
explicit that the reaction force is determined by the absorptive part of
the causal density response, rather than by a phenomenological force
multiplier.

\subsection{The Maxwellian statistical background}

For a three-dimensional Maxwellian mass distribution,
\begin{align}
F_0(\vvec)&=\frac{\rho_0}{(2\pi\sigma^2)^{3/2}}
\exp[-v^2/(2\sigma^2)],\label{eq:Max}\\
f(u)&=\frac{\rho_0}{\sqrt{2\pi}\sigma}
\exp[-u^2/(2\sigma^2)],\label{eq:fmarg}
\end{align}
$f$ is the one-dimensional marginal along any wave-vector direction.
Set $s=\hat\kvec\cdot\Vvec$ and $\delta=\eta k/2$.
The exact imaginary part of Eq.~\eqref{eq:Wchi} is
\begin{align}
\operatorname{Im}\chi_\eta^R
&=\frac{\pi}{\eta k}[f(s+\delta)-f(s-\delta)]
\nonumber\\
&=-\frac{\pi f(s)}{\delta}
 e^{-\delta^2/(2\sigma^2)}
 \sinh(s\delta/\sigma^2).
\label{eq:Imexact}
\end{align}
This expression is exact within the specified bare, linear Wigner
response. It is convenient to choose a Gaussian source
\begin{equation}
W_b(k)=e^{-b^2k^2/2},\qquad b>0,
\label{eq:WGaussian}
\end{equation}
and an infrared cutoff $k_{\min}>0$ describing a finite outer wake scale.
Choosing $\Vvec$ as the polar axis and $x=\hat\kvec\cdot\hat\Vvec$
yields the finite-wavelength drag integral
\begin{equation}
\Fvec_\eta=-4\pi G^2\mu^2\hat\Vvec
\int_{k_{\min}}^\infty\frac{\dd k}{k}e^{-b^2k^2}
\int_{-1}^1\dd x\,x f(Vx)
\exp\!\left[-\frac{\eta^2k^2}{8\sigma^2}\right]
\frac{\sinh\!\left(\eta kVx/(2\sigma^2)\right)}{\eta k/2}.
\label{eq:exactforce}
\end{equation}
For a Maxwellian, the angular integral can actually be carried out
\emph{without a finite-wavelength expansion}. Define
\begin{equation}
X=\frac{V}{\sqrt{2}\sigma},\qquad
 y_k=\frac{\eta k}{2\sqrt{2}\sigma}.
\label{eq:XYwave}
\end{equation}
We obtain a dimensionless spectral friction function
\begin{equation}
\mathcal G_{\rm w}(X,y)=
\frac{\operatorname{erf}(X+y)-\operatorname{erf}(y-X)}{2}
-\frac{e^{-(X-y)^2}-e^{-(X+y)^2}}{2\sqrt\pi\,y}
\qquad(y>0),
\label{eq:waveG}
\end{equation}
with $\mathcal G_{\rm w}(X,0)$ defined by continuity. Hence our
\emph{complete bare linear-response drag} reduces to one wave-number
quadrature,
\begin{equation}
\Fvec_\eta=-\frac{4\pi G^2\mu^2\rho_0}{V^2}\hat\Vvec
\int_{k_{\min}}^\infty\frac{\dd k}{k}e^{-b^2k^2}
\mathcal G_{\rm w}(X,y_k).
\label{eq:exact1d}
\end{equation}
The intermediate angular identity and its normalization are derived in
Appendix~\ref{app:angles}. At fixed $X>0$ the spectral function is
nonnegative; consequently Eq.~\eqref{eq:exact1d} describes a mean drag,
not a sustained thrust, in the specified stationary Maxwellian background.
It is the \emph{same} causal response with either a classical ($y_k\to0$)
or a finite-wave ($y_k>0$) energy transfer spectrum. We emphasize that
``complete'' here refers only to the chosen \emph{bare linear} response,
not to nonlinear gravitational scattering or an arbitrary FDM halo.

\subsection{Chandrasekhar friction with matched cutoffs}

For fixed $k$ and $\eta\to0$,
\begin{equation}
\operatorname{Im}\chi_\eta^R\longrightarrow
\pi f'(s)=-\pi\frac{s}{\sigma^2}f(s).
\label{eq:imclassical}
\end{equation}
With $X$ defined in Eq.~\eqref{eq:XYwave}, let
\begin{equation}
\mathcal G(X)=\operatorname{erf}(X)
-\frac{2X}{\sqrt\pi}e^{-X^2}.
\label{eq:GX}
\end{equation}
The exact spectral function satisfies
\begin{equation}
\mathcal G_{\rm w}(X,y)
=\mathcal G(X)-\frac{4X^3e^{-X^2}}{3\sqrt\pi}\,y^2+O(y^4).
\label{eq:waveGexpand}
\end{equation}
This explicitly isolates the wave physics: each wavenumber probes a
finite velocity interval of width $\eta k$, while the classical
collisionless medium responds to the local velocity gradient.
The Maxwellian angular integral is
\begin{equation}
\int_{-1}^1\dd x\,x f'(Vx)
=-\frac{\rho_0}{V^2}\mathcal G(X).
\label{eq:classangular}
\end{equation}
The Gaussian-regulated Coulomb logarithm is
\begin{equation}
\mathcal L_b=\int_{k_{\min}}^\infty\frac{\dd k}{k}e^{-b^2k^2}
=\frac{1}{2}E_1(z),\qquad z=b^2k_{\min}^2,
\label{eq:Lb}
\end{equation}
where $E_1(z)=\int_z^\infty\dd t\,e^{-t}/t$.
Consequently Eq.~\eqref{eq:masterforce} reduces to
\begin{equation}
\Fvec_{\Ch}^{(b)}=-\frac{4\pi G^2\mu^2\rho_0}{V^2}
\mathcal L_b\mathcal G(X)\hat\Vvec.
\label{eq:Chandrasekhar}
\end{equation}
Thus the collisionless Chandrasekhar result is recovered with the \emph{same}
short-distance Gaussian form factor and infrared cutoff. Replacing
$\mathcal L_b$ by a different arbitrary $\ln\Lambda$ in only one of the
models would not constitute a controlled comparison.

\section{Leading finite-wavelength correction}
\label{sec:correction}

\subsection{Semiclassical expansion of the spectral response}

We now investigate the leading finite-wavelength correction to gravitational dynamical friction and its dependence on the spatial structure of the perturber. The starting point is the exact Wigner susceptibility derived in Sec.~\ref{sec:wigner}, rather than a phenomenological modification of the classical Chandrasekhar force. For a homogeneous Maxwellian background, the angular integration of the bare linear-response drag can be performed analytically, as shown in Sec.~\ref{sec:drag}.

For a general isotropic source characterized by a normalized Fourier form factor $W(k)$, the resulting force can be expressed as
\begin{equation}
\begin{aligned}
\Fvec_\eta
={}&-\frac{4\pi G^2\mu^2\rho_0}{V^2}\hat\Vvec
\int_{k_{\min}}^\infty\frac{\dd k}{k}|W(k)|^2
\mathcal G_{\rm w}(X,y_k),
\end{aligned}
\label{eq:generalforcekernel}
\end{equation}
where $X=V/(\sqrt{2}\sigma)$ and $y_k=\eta k/(2\sqrt{2}\sigma)$, with $\eta=\hbar/m_\chi$. The spectral function $\mathcal G_{\rm w}(X,y_k)$ contains the finite-wavelength dependence of the Maxwellian response, whereas the source structure enters through the nonnegative weight $|W(k)|^2$.

In the semiclassical regime, the spectral function admits the expansion given in Eq.~\eqref{eq:waveGexpand},
\begin{equation}
\mathcal G_{\rm w}(X,y)
=
\mathcal G(X)
-\frac{4X^3e^{-X^2}}{3\sqrt\pi}y^2
+O(y^4),
\label{eq:waveexpVI}
\end{equation}
where
\begin{equation}
\mathcal G(X)
=
\operatorname{erf}(X)
-\frac{2X}{\sqrt\pi}e^{-X^2}.
\label{eq:GVI}
\end{equation}

The absence of a term linear in $y$ reflects the symmetric structure of the underlying Wigner response. The same conclusion follows directly from the expansion of the imaginary part of the retarded susceptibility,
\begin{equation}
\operatorname{Im}\chi_\eta^R(k,s)
=
\pi f'(s)
+\frac{\pi\eta^2k^2}{24}f'''(s)
+O(\eta^4k^4),
\label{eq:Imexp}
\end{equation}
where $s=\hat\kvec\cdot\Vvec$. The two expansions provide complementary algebraic checks of the same response calculation and lead to identical second-order coefficients.

A significant consequence of Eq.~\eqref{eq:waveexpVI} is that different orders in the wavelength expansion probe different spectral moments of the source. The classical contribution contains the radial weight $\dd k/k$, whereas the leading quadratic correction introduces an additional factor of $k^2$. Higher-order corrections involve progressively higher powers of the wave number and are therefore increasingly sensitive to the ultraviolet structure of the perturber.

To make this dependence explicit, we consider a general spherically symmetric finite-size source with a real Fourier form factor satisfying $W(0)=1$. We assume a fixed infrared cutoff $k_{\min}>0$ and introduce the spectral moments
\begin{align}
\mathcal L_W
&=
\int_{k_{\min}}^\infty\frac{\dd k}{k}|W(k)|^2,
\label{eq:LWgeneral}\\
\mathcal J_W
&=
\int_{k_{\min}}^\infty\dd k\,k|W(k)|^2,
\label{eq:JWgeneral}\\
\mathcal K_W
&=
\int_{k_{\min}}^\infty\dd k\,k^3|W(k)|^2.
\label{eq:KWgeneral}
\end{align}

These moments play distinct roles in the calculation. The quantity $\mathcal L_W$ determines the regulated classical Coulomb factor, $\mathcal J_W$ controls the leading finite-wavelength correction, and $\mathcal K_W$ enters the absolute remainder estimate. The separation between these contributions provides a systematic way to identify how the spatial structure of the perturber affects the gravitational drag.

For the asymptotic analysis below, the source profile is held fixed while $\eta\rightarrow0$. The finiteness of the relevant spectral moments is essential for interchanging the wavelength expansion with the wave-number integration. Both the Gaussian and exponential profiles considered below satisfy these requirements.

\subsection{Leading correction for general isotropic perturbers}

Substituting the Wigner expansion \eqref{eq:Imexp} into the spectral-force formula \eqref{eq:masterforce}, we separate the classical contribution from the leading finite-wavelength correction,
\begin{equation}
\Fvec_\eta
=
\Fvec_{\Ch}^{[W]}
+\eta^2\Fvec_2^{[W]}
+\Fvec_{\rm rem}.
\label{eq:forceexp}
\end{equation}

The classical reference force, evaluated with the same source form factor and infrared cutoff, is
\begin{equation}
\Fvec_{\Ch}^{[W]}
=
-\frac{4\pi G^2\mu^2\rho_0}{V^2}
\mathcal L_W\mathcal G(X)\hat\Vvec.
\label{eq:Chgeneral}
\end{equation}

The coefficient of the quadratic contribution can be written as
\begin{align}
\Fvec_2^{[W]}
={}&
\frac{(4\pi G)^2\mu^2\pi}{24(2\pi)^3}
\left[
\int_{k_{\min}}^\infty\dd k\,k|W(k)|^2
\right]
\nonumber\\
&\times
\left[
\int\dd\Omega_\kvec\,
\hat\kvec f'''(\hat\kvec\cdot\Vvec)
\right].
\label{eq:F2start}
\end{align}

For the one-dimensional Maxwellian marginal,
\begin{equation}
f(u)
=
\frac{\rho_0}{\sqrt{2\pi}\sigma}
e^{-u^2/(2\sigma^2)},
\label{eq:fVI}
\end{equation}
the third derivative is
\begin{equation}
f'''(u)
=
\left(
\frac{3u}{\sigma^4}
-\frac{u^3}{\sigma^6}
\right)f(u).
\label{eq:fthird}
\end{equation}

Choosing the polar axis along the perturber velocity, $x=\hat\kvec\cdot\hat\Vvec$, rotational symmetry eliminates the transverse components. The remaining angular integral can be evaluated analytically by integration by parts,
\begin{align}
\int\dd\Omega_\kvec\,
\hat\kvec f'''(Vx)
&=
2\pi\hat\Vvec
\int_{-1}^{1}\dd x\,x f'''(Vx)
\nonumber\\
&=
\frac{4\pi V}{\sigma^4}
f(V)\hat\Vvec.
\label{eq:angularF2}
\end{align}

Substituting this result into Eq.~\eqref{eq:F2start}, we obtain the general quadratic coefficient
\begin{equation}
\Fvec_2^{[W]}
=
\frac{\pi G^2\mu^2\rho_0V}
{3\sqrt{2\pi}\sigma^5}
e^{-X^2}\mathcal J_W\hat\Vvec.
\label{eq:F2general}
\end{equation}

Equation~\eqref{eq:F2general} is the central analytic result of the general-source calculation. It demonstrates that, for a Maxwellian statistical background, the leading finite-wavelength correction depends on the source profile only through the positive spectral moment $\mathcal J_W$. The velocity dependence is determined entirely by the Maxwellian response.

An important physical consequence follows from the sign of this expression. For $V>0$ and a nontrivial isotropic source satisfying $0<\mathcal J_W<\infty$, the quadratic correction points along $+\hat\Vvec$, whereas the classical Chandrasekhar force points along $-\hat\Vvec$. Consequently, the leading finite-wavelength contribution reduces the magnitude of the opposing classical drag within the controlled semiclassical regime.

The profile-independent sign originates from the separation between the angular response of the Maxwellian medium and the radial spectrum of the gravitational source. The Maxwellian angular integral fixes the direction of the leading correction, while the source contributes through the nonnegative spectral weight $k|W(k)|^2$. Changing the isotropic mass profile therefore modifies the magnitude of the correction without changing its sign, provided that the required spectral moments remain finite.

This result should not be interpreted as a universal suppression of dynamical friction. The definite sign follows from the specified Maxwellian distribution, the bare linear-response approximation, and the controlled expansion. A different statistical background or a self-consistent collective gravitational response may lead to different behavior.

To separate the velocity dependence from the source structure, we define an effective squared wave number
\begin{equation}
k_{\rm eff}^2
=
\frac{\mathcal J_W}{\mathcal L_W},
\label{eq:keff}
\end{equation}
and the dimensionless velocity function
\begin{equation}
\mathcal S(X)
=
\frac{X^3e^{-X^2}}
{6\sqrt\pi\,\mathcal G(X)}.
\label{eq:Svelocity}
\end{equation}

The leading finite-wavelength result can then be written in the compact form
\begin{equation}
\begin{aligned}
\Fvec_\eta
=
\Fvec_{\Ch}^{[W]}
\left[
1-\mathcal S(X)
\left(\frac{\eta k_{\rm eff}}{\sigma}\right)^2
\right]
+\Fvec_{\rm rem}.
\end{aligned}
\label{eq:generalresult}
\end{equation}

This expression reveals a factorized structure of the leading relative correction. The function $\mathcal S(X)$ describes the response to the perturber velocity relative to the background dispersion, whereas $k_{\rm eff}$ characterizes the source-weighted spatial scales contributing to the gravitational wake.

The effective wave number is therefore not introduced as an arbitrary cutoff. It emerges from the ratio of the spectral moments appearing in the classical and quadratic contributions. Nevertheless, the smallness of $\eta k_{\rm eff}/\sigma$ alone does not guarantee control of all higher-order terms for an arbitrary source spectrum. The ultraviolet tail must also satisfy the convergence conditions discussed below.

The velocity dependence is particularly transparent in two limiting regimes. For $X\ll1$, the Chandrasekhar function satisfies $\mathcal G(X)=4X^3/(3\sqrt\pi)+O(X^5)$, yielding $\mathcal S(X)\rightarrow1/8$. For $X\gg1$, we obtain $\mathcal S(X)\sim X^3e^{-X^2}/(6\sqrt\pi)$, which is exponentially small.

Thus, the relative quadratic coefficient approaches a finite value in the low-velocity limit and decreases exponentially at large velocities. The absolute gravitational drag nevertheless vanishes as $V\rightarrow0$, and the relative expansion must be interpreted as a limiting relation rather than a ratio evaluated directly at zero velocity.

\subsection{Gaussian profile as an analytic consistency check}

We now specialize the general result to the Gaussian source profile used in Sec.~\ref{sec:drag},
\begin{equation}
W_b(k)
=
e^{-b^2k^2/2},
\qquad b>0.
\label{eq:Wbcorrection}
\end{equation}

Defining $z=b^2k_{\min}^2$, the corresponding spectral moments are
\begin{align}
\mathcal L_b
&=
\frac12 E_1(z),
\label{eq:Lbrepeated}\\
\mathcal J_b
&=
\int_{k_{\min}}^\infty\dd k\,k e^{-b^2k^2}
=
\frac{e^{-z}}{2b^2},
\label{eq:radialF2}\\
\mathcal K_b
&=
\frac{e^{-z}(1+z)}{2b^4}.
\label{eq:KbGaussian}
\end{align}

Substituting $\mathcal J_b$ into Eq.~\eqref{eq:F2general}, we recover the Gaussian coefficient obtained from the direct Maxwellian response calculation,
\begin{equation}
\Fvec_2^{(b)}
=
\frac{\pi G^2\mu^2\rho_0}
{6\sqrt{2\pi}\,b^2\sigma^5}
V e^{-X^2-z}\hat\Vvec.
\label{eq:F2result}
\end{equation}

It is convenient to define the dimensionless wave parameter
\begin{equation}
\epsilon_{\rm w}
=
\frac{\eta}{b\sigma}
=
\frac{\hbar}{m_\chi b\sigma},
\label{eq:epsilon}
\end{equation}
which compares the characteristic wave-induced velocity shift at the source scale with the background velocity dispersion.

Introducing the Gaussian correction coefficient
\begin{equation}
\mathcal A(X,z)
=
\frac{X^3e^{-X^2-z}}
{12\sqrt\pi\,\mathcal L_b\mathcal G(X)},
\label{eq:Acoef}
\end{equation}
the general-source result reduces to
\begin{equation}
\begin{aligned}
\Fvec_\eta
={}&
-\frac{4\pi G^2\mu^2\rho_0}{V^2}
\mathcal L_b\mathcal G(X)
\\
&\times
\left[
1-\mathcal A(X,z)\epsilon_{\rm w}^2
\right]\hat\Vvec
+\Fvec_{\rm rem}.
\end{aligned}
\label{eq:mainresult}
\end{equation}

The agreement between this expression and the direct expansion of the exact Maxwellian angular kernel provides an important consistency check of the normalization and sign of the quadratic coefficient. In particular, the Gaussian correction is recovered without introducing any additional assumptions beyond the source form factor.

For small and large velocity ratios, respectively, the coefficient satisfies
\begin{align}
\mathcal A(X,z)
&\longrightarrow
\frac{e^{-z}}{16\mathcal L_b},
\qquad X\ll1,
\label{eq:lowX}\\
\mathcal A(X,z)
&\sim
\frac{X^3e^{-X^2-z}}
{12\sqrt\pi\,\mathcal L_b},
\qquad X\gg1.
\label{eq:highX}
\end{align}

The positive coefficient $\Fvec_2^{(b)}$ reduces the magnitude of the classical drag; it does not represent an additional propulsive force. For wave parameters outside the controlled semiclassical regime, the unexpanded result in Eq.~\eqref{eq:exact1d} should be used instead of extrapolating the quadratic approximation.

\subsection{A non-Gaussian finite-source example}

To demonstrate that the leading correction is not restricted to Gaussian regulators, we consider a normalized spherical exponential mass distribution,
\begin{equation}
\rho_p(r)
=
\frac{\mu}{8\pi a^3}e^{-r/a},
\qquad a>0.
\label{eq:exponentialprofile}
\end{equation}

The corresponding Fourier form factor is
\begin{equation}
W_a(k)
=
\frac{1}{(1+a^2k^2)^2}.
\label{eq:exponentialW}
\end{equation}

Unlike the Gaussian profile, this source produces an algebraically decaying form factor at large wave numbers. It therefore provides a useful independent example for examining the dependence of the semiclassical correction on the ultraviolet source structure.

Introducing $q=a^2k_{\min}^2$, the required spectral moments can be evaluated analytically:
\begin{align}
\mathcal L_a
&=
\frac12
\left[
\ln\frac{1+q}{q}
-
\sum_{j=1}^{3}
\frac{1}{j(1+q)^j}
\right],
\label{eq:exponentialL}\\
\mathcal J_a
&=
\frac{1}{6a^2(1+q)^3},
\label{eq:exponentialJ}\\
\mathcal K_a
&=
\frac{1}{4a^4(1+q)^2}
-
\frac{1}{6a^4(1+q)^3}.
\label{eq:exponentialK}
\end{align}

The leading correction follows immediately from Eq.~\eqref{eq:F2general},
\begin{equation}
\Fvec_2^{(a)}
=
\frac{\pi G^2\mu^2\rho_0V}
{3\sqrt{2\pi}\sigma^5}
e^{-X^2}\mathcal J_a\hat\Vvec.
\label{eq:F2exponential}
\end{equation}

Since $\mathcal J_a>0$ for $a>0$, the quadratic correction again points along the perturber velocity and reduces the magnitude of the classical drag in the controlled semiclassical regime.

The Gaussian and exponential sources therefore exhibit the same leading correction direction despite their different Fourier spectra. Their correction magnitudes are different because the corresponding spectral moments are different. These examples illustrate the general result established in Eq.~\eqref{eq:F2general}, rather than constituting independent assumptions about the sign of the wave correction.

\subsection{Controlled remainder and range of validity}

The preceding results follow from a small-$\eta$ expansion of the Wigner susceptibility. To establish that the expansion remains controlled after integration over wave number, we now estimate the higher-order remainder of the complete force.

For the one-dimensional Maxwellian marginal, the fifth velocity derivative is bounded. Applying the centered Taylor remainder formula to the Wigner finite difference gives
\begin{equation}
\left|R_{\operatorname{Im}\chi}\right|
\leq
\frac{\pi\eta^4k^4}{1920}
\sup_{u\in\mathbb R}|f^{(5)}(u)|.
\label{eq:chibound}
\end{equation}

Substituting this inequality into the spectral-force integral and using the isotropy of the source, we obtain the profile-dependent absolute estimate
\begin{equation}
\|\Fvec_{\rm rem}\|
\leq
\frac{\pi G^2\mu^2\eta^4}{240}
\mathcal K_W
\sup_{u\in\mathbb R}|f^{(5)}(u)|.
\label{eq:generalbound}
\end{equation}

The appearance of $\mathcal K_W$ is physically significant. While $\mathcal J_W$ measures the source contribution to the leading quadratic correction, the higher spectral moment $\mathcal K_W$ controls the sensitivity of the neglected terms to the ultraviolet part of the gravitational source spectrum.

For a Gaussian source, substituting Eq.~\eqref{eq:KbGaussian} yields
\begin{equation}
\begin{aligned}
\|\Fvec_{\rm rem}\|
\leq{}&
\frac{\pi G^2\mu^2\eta^4}{480b^4}
e^{-z}(1+z)
\\
&\times
\sup_{u\in\mathbb R}|f^{(5)}(u)|.
\end{aligned}
\label{eq:Fbound}
\end{equation}

This expression reproduces the fixed-radius Gaussian remainder bound. Thus, for a fixed isotropic source with finite positive moments $\mathcal L_W$, $\mathcal J_W$, and $\mathcal K_W$, the gravitational drag admits an asymptotic expansion with an absolute remainder of order $O(\eta^4)$.

The finiteness of $\mathcal K_W$ is a sufficient condition for this force-level control. It ensures that the ultraviolet part of the source spectrum does not invalidate the integrated Taylor estimate, even though the pointwise expansion at fixed wave number is not uniform over arbitrarily large $k$.

It is important, however, to distinguish an absolute remainder estimate from a relative error bound. As the perturber velocity approaches zero, the classical drag also vanishes, and a finite absolute bound does not by itself guarantee a uniformly small relative error. Similarly, shrinking the source size generally increases the higher spectral moments, so the expansion need not remain uniform in the formal point-source limit.

The leading reduction of dynamical friction established above is therefore a conditional result. It applies to a stationary homogeneous Maxwellian background, a normalized isotropic finite-size source, a weak bare response without collective self-gravitating dressing, and a controlled semiclassical expansion. The present calculation does not establish a universal correction sign for anisotropic or non-Maxwellian backgrounds, nonlinear gravitational scattering, or self-consistent inhomogeneous dark matter halos.

The nonuniform point-source limit and the physical significance of the relevant gravitational scales are discussed in Sec.~\ref{sec:limits}.

\section{Physical regimes and the point-source limit}
\label{sec:limits}

The validity of the finite-wavelength expansion depends on both the wave properties of the FDM background and the spatial structure of the gravitational perturber. For a general isotropic source, the classical contribution, the leading quadratic correction, and the higher-order remainder are controlled by the spectral moments $\mathcal L_W$, $\mathcal J_W$, and $\mathcal K_W$, respectively. As established in Sec.~\ref{sec:correction}, a finite $\mathcal K_W$ provides a sufficient condition for an absolute $O(\eta^4)$ remainder at fixed source profile. However, this result does not imply that the expansion remains uniform when the source is continuously contracted toward a point mass.

To clarify this distinction, consider a family of isotropic source profiles characterized by a length scale $b$, with $W_b(k)=W_1(bk)$. Introducing the dimensionless wave number $p=bk$, the moments governing the quadratic correction and its remainder become
\begin{align}
\mathcal J_b
&=
\frac{1}{b^2}
\int_{bk_{\min}}^\infty
\dd p\,p|W_1(p)|^2,
\\
\mathcal K_b
&=
\frac{1}{b^4}
\int_{bk_{\min}}^\infty
\dd p\,p^3|W_1(p)|^2.
\end{align}
For source profiles whose corresponding integrals approach finite nonzero limits as $bk_{\min}\to0$, these expressions scale as $b^{-2}$ and $b^{-4}$, respectively. The increasing sensitivity to high wave numbers shows that the fixed-source semiclassical expansion is not generally uniform in the point-source limit.

For the Gaussian form factor $W_b(k)=e^{-b^2k^2/2}$, the characteristic expansion parameter is
\begin{equation}
\epsilon_{\rm w}
=
\frac{\eta}{b\sigma}
=
\frac{\hbar}{m_\chi b\sigma}.
\end{equation}
The Wigner expansion requires $\eta k/\sigma\ll1$ over the wave numbers relevant to the correction. Since the Gaussian profile suppresses contributions with $k\gg b^{-1}$, the condition $\epsilon_{\rm w}\ll1$ provides a convenient sufficient scale hierarchy for the fixed-radius expansion. The leading coefficient in Eq.~\eqref{eq:F2result} scales as $b^{-2}$, while the remainder bound in Eq.~\eqref{eq:Fbound} scales as $b^{-4}$ when $bk_{\min}\ll1$. Thus, the quadratic approximation cannot be extrapolated to arbitrarily small source radii at fixed nonzero $\eta$.

The breakdown of the truncated expansion should not be confused with a divergence of the complete finite-wavelength force. For a homogeneous Maxwellian statistical background, the imaginary part of the unexpanded Wigner susceptibility is given by Eq.~\eqref{eq:Imexact}. At fixed $\eta>0$, $\sigma>0$, and finite source velocity, its large-$k$ behavior contains the Gaussian suppression
\begin{equation}
\operatorname{Im}\chi_\eta^R
\propto
\exp\left[
-\frac{\eta^2k^2}{8\sigma^2}
+O(k)
\right],
\qquad k\rightarrow\infty.
\end{equation}
Consequently, with a fixed positive infrared cutoff, the ultraviolet part of the formal bare linear-response force remains convergent even when the source form factor approaches the point-source value $W(k)=1$. This mathematical convergence does not establish the physical validity of linear response arbitrarily close to a point mass. Nonlinear gravitational focusing and large density perturbations may become important before the formal point-source limit is reached.

The distinction between the complete response and its semiclassical expansion can be illustrated explicitly using the Maxwellian angular kernel. In the formal point-source case $W(k)=1$, introduce $y=\eta k/(2\sigma)$ and define
\begin{equation}
H_X(y)
=
\frac{\mathcal G_{\rm w}(X,y/\sqrt2)}
{\mathcal G(X)},
\qquad X>0.
\label{eq:HX}
\end{equation}
For fixed $X>0$, this function approaches unity as $y\rightarrow0$ and decreases rapidly at large $y$. More precisely, its small-$y$ expansion is
\begin{equation}
H_X(y)
=
1-
\frac{2X^3e^{-X^2}}
{3\sqrt\pi\,\mathcal G(X)}y^2
+O(y^4).
\end{equation}

The formal point-source drag can therefore be written as
\begin{equation}
\begin{aligned}
\Fvec_\eta^{\rm point}
={}&
-\frac{4\pi G^2\mu^2\rho_0}{V^2}
\mathcal G(X)\hat\Vvec
\\
&\times
\int_{\eta k_{\min}/(2\sigma)}^\infty
\frac{\dd y}{y}H_X(y).
\end{aligned}
\label{eq:Fpoint}
\end{equation}

For small $\varepsilon=\eta k_{\min}/(2\sigma)$, the integral admits the asymptotic form
\begin{equation}
\int_\varepsilon^\infty
\frac{\dd y}{y}H_X(y)
=
\ln\frac1\varepsilon
+\mathcal B(X)
+O(\varepsilon^2),
\qquad
\varepsilon=\frac{\eta k_{\min}}{2\sigma},
\label{eq:logpoint}
\end{equation}
where
\begin{equation}
\begin{aligned}
\mathcal B(X)
={}&
\int_0^1\frac{\dd y}{y}
[H_X(y)-1]
\\
&+
\int_1^\infty
\frac{\dd y}{y}H_X(y).
\end{aligned}
\label{eq:BX}
\end{equation}
Both integrals defining $\mathcal B(X)$ are absolutely convergent for fixed $X>0$. The logarithmic term arises because $H_X(y)\to1$ at small $y$, while the finite-wavelength response suppresses sufficiently large wave numbers.

Within the idealized bare linear-response model, this result illustrates how the wave scale provides an effective ultraviolet regularization of the point-source spectrum. Nevertheless, the logarithm should not be interpreted as a universal physical Coulomb logarithm. In particular, the classical limit at fixed finite source size and the point-source limit require different ultraviolet treatments. The nonuniformity of the semiclassical expansion prevents the point-source result from being obtained by simply taking $b\to0$ in the quadratic correction derived for a finite source.

For a realistic gravitational perturber, the relevant short-distance physics may involve the physical source radius, nonlinear scattering, and gravitational focusing. The characteristic strong-deflection scale depends on the relative velocity between the perturber and the background particles,
\begin{equation}
b_{90}(v_{\rm rel})
\sim
\frac{G\mu}{v_{\rm rel}^2}.
\end{equation}
In a statistical Maxwellian medium, the relative velocity is distributed rather than fixed to the bulk perturber speed $V$. Therefore, the physical matching scale cannot in general be determined from $G\mu/V^2$ alone.

The controlled Gaussian-source regime considered in this work may be summarized by
\begin{equation}
\epsilon_{\rm w}
=
\frac{\hbar}{m_\chi b\sigma}
\ll1,
\qquad
\frac{G\mu}{b\sigma^2}\ll1,
\qquad
V,\sigma\ll c.
\label{eq:valid}
\end{equation}
The first condition controls the finite-wavelength expansion at the characteristic source scale, while the second provides a convenient weak-source estimate relative to the background velocity dispersion. These inequalities do not replace a direct verification of the linear-response condition $|\delta\rho|/\rho_0\ll1$.

In addition, the bare-response approximation requires collective self-gravitating corrections to remain small over the modes contributing significantly to the drag. Away from collective resonances, a sufficient condition is
\begin{equation}
\left|
\frac{4\pi G}{k^2}
\chi_\eta^R(\kvec,\omega)
\right|
\ll1,
\end{equation}
with the susceptibility evaluated on the relevant source frequencies. Modes affected by collective instabilities require separate treatment and cannot be described by the stationary bare-response approximation.

The infrared cutoff $k_{\min}$ must also be associated with a specified physical outer scale or interaction history when applying the model to an astrophysical system. Our results therefore define a controlled finite-source response problem rather than a universal prescription for gravitational friction in arbitrary FDM halos. In regimes where the semiclassical expansion is not justified, the unexpanded Wigner response should be used, together with an independent assessment of nonlinear and collective gravitational effects.

\section{Discussion and conclusions}
\label{sec:conclusion}

In this work, we have investigated finite-wavelength corrections to gravitational dynamical friction in fuzzy dark matter (FDM) within a causal linear-response framework. Starting from the Schr\"odinger--Poisson description, we connected the wave dynamics of the medium to its induced density response and the gravitational reaction force acting on a moving perturber. The Schwinger--Keldysh influence functional establishes the causal structure of this backreaction, while the statistical Wigner formulation incorporates the finite-wavelength properties of a background with nonzero velocity dispersion. Together, these approaches provide a consistent description of gravitational drag without introducing a phenomenological correction to the classical Chandrasekhar formula.

The underlying physical mechanism is the gravitational backreaction of the density wake generated by a moving source. The external gravitational potential excites disturbances in the FDM medium, and the induced density distribution produces an additional gravitational field acting on the perturber. In general, the reaction force depends on the past history of the source through the retarded density susceptibility. For uniform motion in a stationary homogeneous background, this response admits a spectral representation in which the mean drag is determined by the absorptive part of the retarded susceptibility. The Wigner response differs from its classical counterpart through a finite difference of the background velocity distribution, which reduces to the velocity derivative of classical Vlasov theory in the semiclassical limit.

For a homogeneous Maxwellian statistical background, we obtained an analytic angular integration of the complete bare linear-response force, reducing the finite-wavelength drag to a single integral over wave number. This representation retains the full wavelength dependence of the specified linear response and provides a reference for evaluating semiclassical approximations. When the wave and classical calculations employ identical source profiles and infrared cutoffs, the classical Chandrasekhar expression is recovered. Such matching is essential for separating genuine finite-wavelength effects from differences associated with the regularization of gravitational interactions.

The principal analytic result is the leading finite-wavelength correction for a general class of isotropic finite-size perturbers. For a normalized source with Fourier form factor $W(k)$, the classical drag is governed by the regulated Coulomb factor $\mathcal L_W$, whereas the quadratic wave correction depends on the spectral moment $\mathcal J_W$. Defining $k_{\rm eff}^2=\mathcal J_W/\mathcal L_W$, we find that the leading relative correction factorizes into a Maxwellian velocity-dependent function and a source-dependent effective wave-number scale. This structure separates the influence of the statistical background from that of the spatial distribution of the gravitational source.

An important consequence is that the leading quadratic correction has a definite sign within the adopted Maxwellian model. For a nontrivial isotropic finite-size source with finite positive spectral moments, the correction points along the perturber velocity and therefore reduces the magnitude of the opposing classical drag in the controlled semiclassical regime. The detailed radial profile changes the correction coefficient through its spectral moments, but not its sign. This property follows from the separation between the Maxwellian angular response and the nonnegative radial source weight. It is not, however, a universal suppression law for arbitrary FDM backgrounds, since the result relies on the homogeneous Maxwellian distribution, the bare linear-response approximation, and the validity of the finite-wavelength expansion.

We illustrated the general result using Gaussian and exponential source profiles. Both admit explicit analytic spectral moments and recover the same leading correction direction despite their different spatial structures. The relative correction approaches a finite coefficient in the low-velocity limit, although the absolute drag vanishes as the perturber velocity tends to zero. At large velocities relative to the Maxwellian dispersion, the leading relative correction becomes exponentially small. These properties demonstrate that finite-wavelength modifications depend on the source velocity, background velocity dispersion, and characteristic spatial scales of the perturber, rather than on the FDM boson mass alone.

The validity of the semiclassical expansion requires particular attention to the ultraviolet structure of the source. While $\mathcal J_W$ determines the leading quadratic coefficient, the higher spectral moment $\mathcal K_W$ controls the absolute fourth-order remainder. For a fixed isotropic source satisfying the required convergence conditions, the drag admits an expansion with an absolute $O(\eta^4)$ remainder, where $\eta=\hbar/m_\chi$. However, this estimate is not uniform when the source size approaches zero. In the Gaussian example, the leading coefficient scales as $b^{-2}$, while the remainder bound scales as $b^{-4}$. This behavior reflects the breakdown of the fixed-source expansion in the formal point-source limit, rather than a necessary divergence of the unexpanded finite-wavelength force. A physically meaningful point-source treatment must additionally account for gravitational focusing, nonlinear perturbations, and the relevant short-distance scales.

Several assumptions restrict the applicability of these conclusions. Our calculations employ a homogeneous isotropic Maxwellian statistical background and neglect collective self-gravitating dressing of the induced wake. Including collective gravity may modify the susceptibility and introduce Jeans-unstable modes, while realistic FDM halos can exhibit inhomogeneity, coherent interference, solitonic structures, and anisotropic velocity distributions. Furthermore, although the influence functional contains both a retarded response kernel and a symmetric density-fluctuation kernel, we have focused on the deterministic mean force rather than stochastic gravitational effects. Extending the analysis to more general statistical backgrounds and self-consistent gravitational environments would be necessary before applying the present coefficients to specific astrophysical systems.

The field-theoretic and wave-response descriptions of FDM dynamical friction have important precedents \cite{Lancaster2020,Buehler2023,Proukakis2023,Proukakis2025}. The present analysis combines these established theoretical tools with a matched classical comparison and a controlled finite-source asymptotic expansion. Its main result is the spectral-moment structure of the leading finite-wavelength correction for isotropic perturbers in a Maxwellian background, together with the associated sign property and remainder estimate. These results should be understood within the assumptions of the specified model, and their precise relationship to previous finite-source calculations remains important when assessing their broader significance.

In conclusion, finite-wavelength effects modify dynamical friction through the dispersive and statistical response of the FDM medium to a moving gravitational source. By deriving the leading correction for a general class of isotropic finite-size perturbers, recovering the matched Chandrasekhar limit, and establishing explicit conditions for the validity of the semiclassical expansion, we clarify how the spatial structure of a perturber and the statistical properties of its environment jointly determine the gravitational drag. The resulting formulation provides a basis for extending finite-wavelength response calculations to more realistic wave-like dark matter systems.

\acknowledgments

The authors gratefully acknowledge support from the project
``Integrated Electronics Technology for Inertial Sensors''
(No.~2024YFC2207003), led by Meilin Liu.
This work is also supported by the China--Brazil Belt and Road Joint
Laboratory on Radio Astronomy Technology and by the National Key Research
and Development Program of China under the project
``Strategic Science and Technology Innovation Cooperation''
(No.~2025YFE0212600), also led by Meilin Liu.

\section*{Funding}

This work was supported by the National Key Research and Development
Program of China under Grant Nos.~2024YFC2207003 and 2025YFE0212600,
as well as related programs of the China--Brazil Belt and Road Joint
Laboratory on Radio Astronomy Technology.

\section*{Data Availability Statement}

This work is theoretical and analytical in nature.
No research datasets were generated or analyzed
during the present study.

\section*{Code Availability Statement}

The principal results of this work were obtained
analytically. Any computational checks were used
solely to verify the analytical expressions, and
no dedicated research software is required to
reproduce the theoretical derivations.

\appendix
\section{Quadratic Madelung action and the cold-wave response}
\label{app:kernel}

Introduce $\rho=\rho_0+n$ and $\vartheta=\vartheta_0+\theta$ in
Eq.~\eqref{eq:SPaction}. Up to second order, and after subtracting the
background equilibrium, the fluctuation action is
\begin{equation}
S^{(2)}=\int\dd t\,\dd^3x\left[-nD_0\theta
-\frac{\rho_0}{2}|\bm\nabla\theta|^2
-\frac{\eta^2}{8\rho_0}|\bm\nabla n|^2
-n(\phip+\varphi_1)
-\frac{|\bm\nabla\varphi_1|^2}{8\pi G}\right].
\label{eq:S2app}
\end{equation}
Variation in $\theta$ gives $D_0n+\rho_0\nabla^2\theta=0$.
Variation in $\varphi_1$ gives $\nabla^2\varphi_1=4\pi Gn$.
Eliminating both fields, the density-only quadratic kernel in Fourier
space has the form
\begin{equation}
\mathcal K(\kvec,\omega)=\frac{
[\omega-\kvec\cdot\mathbf u_0]^2
-\eta^2k^4/4+4\pi G\rho_0}{\rho_0k^2}.
\label{eq:Kernel}
\end{equation}
The algebraic inverse is $\mathcal K^{-1}=\chi_{\rm cold}$ without yet
specifying the causal prescription. The CTP formulation selects the
retarded inverse, as used in Eq.~\eqref{eq:coldchi}. For the statistical
background the density susceptibility instead follows from the Gaussian
field two-point function. In the free-field limit,
$\hat\rho=m_\chi\hat\Psi^\dagger\hat\Psi$ and the Kubo commutator reduces to
the difference of two mode occupations, as displayed in
Eq.~\eqref{eq:occupationbubble}; shifting to the midpoint momentum and
using the mass-normalization of $F_0$ then produces Eq.~\eqref{eq:Wchi}.
Thus the classical-field and occupation-number calculations agree on the
\emph{retarded} response, but the cold Madelung kernel must not be
identified with an arbitrary Maxwellian ensemble.

\section{Exact angular reduction and finite-wavelength coefficient}
\label{app:angles}

For a Gaussian one-dimensional marginal and $\delta=\eta k/2$, define
\begin{equation}
I(V,\delta)=\int_{-1}^1\dd x\,x
\left[f(Vx+\delta)-f(Vx-\delta)\right].
\label{eq:Iangle}
\end{equation}
By changing $x\to-x$ in the second contribution and then
$u=Vx+\delta$ in the first one, one finds
\begin{equation}
I(V,\delta)=\frac{2}{V^2}
\left\{\sigma^2[f(V-\delta)-f(V+\delta)]
-\delta\int_{\delta-V}^{\delta+V}\dd u\,f(u)\right\}.
\label{eq:Iangleeval}
\end{equation}
The remaining Gaussian integral is an error-function difference.
The identity
\begin{equation}
-\frac{V^2}{\rho_0\eta k}I(V,\delta)
=\mathcal G_{\rm w}\left(\frac{V}{\sqrt2\sigma},
\frac{\delta}{\sqrt2\sigma}\right)
\label{eq:Gwaveidentity}
\end{equation}
leads immediately to Eqs.~\eqref{eq:waveG} and~\eqref{eq:exact1d}.
For $x>0$ and $\delta>0$, monotonicity of the even Maxwellian in $|u|$
ensures $f(Vx+\delta)\le f(Vx-\delta)$. The corresponding negative-$x$
contribution has the same sign after multiplication by $x$, proving
$\mathcal G_{\rm w}(X,y)\ge0$ for $X,y>0$. Expanding its two terms at
$y=0$ yields Eq.~\eqref{eq:waveGexpand}.

As an independent derivative-based check, the projected Maxwellian obeys
\begin{equation}
f'(u)=-\frac{u}{\sigma^2}f(u),\qquad
f'''(u)=\left(\frac{3u}{\sigma^4}-\frac{u^3}{\sigma^6}\right)f(u).
\end{equation}
For the third derivative, choosing $x=\cos\theta$ gives
\begin{align}
\int_{-1}^1 x f'''(Vx)\,\dd x
&=\frac{1}{V^2}\int_{-V}^{V}u f'''(u)\,\dd u\\
&=\frac{2}{V^2}[Vf''(V)-f'(V)]\\
&=\frac{2V}{\sigma^4}f(V).
\end{align}
Multiplication by $2\pi\hat\Vvec$ gives Eq.~\eqref{eq:angularF2}.
For the radial integral, with $z=b^2k_{\min}^2$,
\begin{equation}
\int_{k_{\min}}^\infty k e^{-b^2k^2}\,\dd k
=\frac{e^{-z}}{2b^2},
\end{equation}
which completes the coefficient in Eq.~\eqref{eq:F2result}.
Its normalization may be checked independently, e.g., for
$G=\mu=\rho_0=V=\sigma=b=1$ and $k_{\min}=0.1$, for which
$F_{\Ch}^{(b)}\simeq-5.04244848$ and the $\eta^2$ coefficient is
$F_2^{(b)}\simeq+0.12543493$ in the direction of motion. These numbers
serve only as an algebraic verification, not as astrophysical results.

\section{Bound on the finite-source expansion remainder}
\label{app:bound}

For any smooth $f$ with bounded fifth derivative, centered Taylor
expansion gives
\begin{equation}
\left|\frac{f(s+\delta)-f(s-\delta)}{2\delta}
-f'(s)-\frac{\delta^2}{6}f'''(s)\right|
\leq\frac{\delta^4}{120}\sup_u|f^{(5)}(u)|.
\label{eq:Taylorbound}
\end{equation}
Using $\delta=\eta k/2$, the factor $\pi$ from the retarded imaginary
part, and $\int\dd\Omega_\kvec\leq4\pi$ in absolute value, one finds
\begin{equation}
\|\Fvec_{\rm rem}\|\leq\frac{\pi G^2\mu^2\eta^4}{240}
\sup|f^{(5)}|\int_{k_{\min}}^\infty\dd k\,k^3e^{-b^2k^2}.
\end{equation}
The integral evaluates to
$e^{-z}(1+z)/(2b^4)$, yielding Eq.~\eqref{eq:Fbound}.
The estimate holds at each fixed $b>0$ and is not a uniform bound in
the zero-size limit.

\end{document}